# Magic ratios for connectivity-driven electrical conductance of graphene-like molecules

Yan Geng,[1,‡] Sara Sangtarash,*,[2,‡] Cancan Huang,[1,‡] Hatef Sadeghi,[2,‡] Yongchun Fu,[1] Wenjing Hong,*,[1] Thomas Wandlowski,[1] Silvio Decurtins,[1] Colin J. Lambert,*,[2] and Shi-Xia Liu*,[1]

[1] Department of Chemistry and Biochemistry, University of Bern, Freiestrasse 3, CH-3012 Bern, Switzerland
[2] Lancaster Quantum Technology Centre, Department of Physics, Lancaster University, LA14YB Lancaster, UK

**ABSTRACT:** Experiments using a mechanically-controlled break junction and calculations based on density functional theory demonstrate a new magic ratio rule (MRR), which captures the contribution of connectivity to the electrical conductance of graphene-like aromatic molecules. When one electrode is connected to a site i and the other is connected to a site i' of a particular molecule, we assign the molecule a "magic integer" $M_{ii'}$. Two molecules with the same aromatic core, but different pairs of electrode connection sites (i,i' and j,j' respectively) possess different magic integers $M_{ii'}$ and $M_{jj'}$. Based on connectivity alone, we predict that when the coupling to electrodes is weak and the Fermi energy of the electrodes lies close to the centre of the HOMO-LUMO gap, the ratio of their conductances is equal to $(M_{ii'}/M_{jj'})^2$. The MRR is exact for a tight binding representation of a molecule and a qualitative guide for real molecules.

**INTRODUCTION** Charge transport through polycyclic aromatic hydrocarbons (PAHs) has attracted intensive attention in recent years[1,2], partly due to their role in the design and development of molecular electronic devices[3-6]. Since PAHs are well-defined and defect free, they also provide model systems for understanding transport in graphene, treated as an infinite alternant PAH, and graphene-based nanostructures[7-9]. When a single molecule is connected to metallic electrodes, electrons passing through the molecule from one electrode to the other can remain phase coherent, even at room temperature[10,11]. This has led to a great deal of discussion about the role of quantum interference (QI) in determining the electrical conductance of single molecules[12-21], culminating in a series of recent experiments revealing room-temperature signatures of QI[22-30].

Both experiment and theory have focused primarily on elucidating the conditions for the appearance of constructive or destructive interference. In the simplest case, where electrons are injected at the Fermi energy $E_F$ of the electrodes, constructive QI arises when $E_F$ coincides with a delocalized energy level $E_n$ of the molecule. Similarly a simple form of destructive QI occurs when $E_F$ coincides with the energy $E_b$ of a bound state located on a pendant moiety[31,32]. In practice, unless energy levels are tuned by electrostatic, electrochemical or mechanical gating, molecules located within a junction rarely exhibit these types of QI, because $E_F$ is usually located in the HOMO-LUMO (H-L) gap. For this reason, discussions have often focussed on conditions for destructive or constructive QI when $E_F$ is located at the centre of the H-L gap. For the purpose of identifying conditions for destructive QI within the delocalised π-system, a useful conceptual approach is to represent molecules by lattices of connected sites (C(sp$^2$) atoms), such as those shown in Fig. 1, in which 1a represents a benzene ring, 1b represents naphthalene, 1c represents anthracene and 1d represents anthanthrene. Such abstractions highlight the role of connectivity in determining the presence or absence of destructive QI. For example, the lattices of Fig. 1 are bipartite, being composed of equal numbers of 'primed' and 'unprimed' sites, such that primed sites (labelled by primed integers such as *1',2',3'*) are connected to unprimed sites only (labelled by non-primed integers such as *1,2,3*) and vice versa. It is well known[33-38], (see also the Mathematical Methods below) that if electrodes are connected to two sites which are both primed or both unprimed, then destructive interference occurs and the contribution from π-orbitals to the electrical conductance $G$ vanishes. For a phenyl ring this corresponds to the well-known case of meta-coupled electrodes[31], but more generally it holds for any bipartite lattice.

Studies of such lattices have yielded a variety of simple rules for the appearance of destructive QI[21,32-37], for which the π-orbital contribution to $G$ vanishes. The aim of the present paper is to elucidate a simple rule for determining the non-zero values of electrical conductance arising from constructive QI in aromatic molecules. At first sight, this task seems rather daunting, because there is only one conductance (*i.e.* $G = 0$) when QI is destructive, whereas there are many possible non-zero values of $G$ when QI is constructive. Furthermore, the non-zero values of conductances in the presence of constructive QI depend on the strength and detailed nature of the contacts to electrodes.

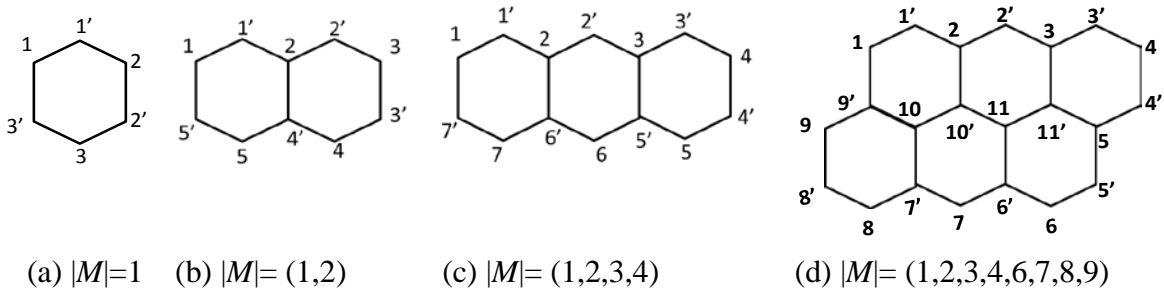

(a) |M|=1    (b) |M|= (1,2)    (c) |M|= (1,2,3,4)    (d) |M|= (1,2,3,4,6,7,8,9)

**Figure 1** Four examples of bipartite lattices, with the magnitude of their magic numbers shown underneath each lattice. (a) represents benzene, (b) naphthalene, (c) anthracene and (d) anthanthrene.

Remarkably, in what follows, we demonstrate a "magic ratio rule" based on tables of quantum numbers $M_{ii'}$, which capture the contribution of connectivity to the electrical conductance of graphene-like aromatic molecules, or molecules with graphene-like cores, when one electrode is connected to an 'unprimed' site $i$ and the other is connected to a 'primed' site $i'$. In particle physics, quantum numbers such as 'charm' and 'colour' are assigned to elementary particles. In the case of lattices such as those in Fig. 1, we refer to these new quantum numbers as 'magic integers $M_{ii'}$'. For each of the molecules shown in Fig. 1, the allowed values of $|M_{ii'}|$ are shown beneath each lattice. Clearly the spectrum of magic integers increases with size of the aromatic core. The precise values of $M_{ii'}$ are not trivial, since for example $M_{ii'} = 5$ is missing from the set of anthanthrene integers.

**RESULTS AND DISCUSSION**

Magic integers (MIs) capture the complexity of interference patterns created by electrons at the centre of HOMO-LUMO gap and allow the prediction of conductance ratios *via* the following 'magic ratio rule' (MRR), which states that "the ratio of conductances of two molecules is equal to the squares of the ratios of their magic integers." Clearly, when comparing conductances of the same aromatic core, but different contacts, the signs of the MIs are irrelevant. This rule is derived in the Mathematical Methods section. To each lattice such as those in Fig. 1, the quantum numbers $M_{ii'}$ form a table of MIs, which we refer to as M-tables. As shown in the SI, for the benzene ring 1a, this is a 3x3 table, with all entries equal to +/-1, so that $|M_{ii'}| = 1$ is the only possibility and therefore as expected, para (*i.e.* 3,1′) or ortho (3,2′ or 3,3′) connectivities yield the same electrical conductances. For the naphthalene lattice 1b, the 5x5 M-table is shown in Table 1. As expected from symmetry, this table shows the conductances associated with contact sites 1,1′ and 5,5′ are equal and proportional to $(2)^2 = 4$. It also shows that the conductance with contact sites 4,2′ or 4,3′ would take the same value, which is a less obvious result.

|   | 1′ | 2′ | 3′ | 4′ | 5′ |
|---|----|----|----|----|----|
| 1 | -2 | 1  | -1 | 1  | -1 |
| 2 | -1 | -1 | 1  | -1 | 1  |
| 3 | 1  | -2 | -1 | 1  | -1 |
| 4 | -1 | 2  | -2 | -1 | 1  |
| 5 | 2  | -1 | 1  | -1 | -2 |

**Table 1** The M-table of MIs $M_{ii'}$ for the naphthalene lattice of Fig. 1b.

|    | 1′ | 2′ | 3′ | 4′ | 5′ | 6′ | 7′ | 8′ | 9′ | 10′ | 11′ |
|----|----|----|----|----|----|----|----|----|----|-----|-----|
| 1  | -9 | 7  | -4 | 4  | -1 | 1  | -1 | 1  | -1 | 2   | -3  |
| 2  | -1 | -7 | 4  | -4 | 1  | -1 | 1  | -1 | 1  | -2  | 3   |
| 3  | 1  | -3 | -4 | 4  | -1 | 1  | -1 | 1  | -1 | 2   | -3  |
| 4  | -1 | 3  | -6 | -4 | 1  | -1 | 1  | -1 | 1  | -2  | 3   |
| 5  | 1  | -3 | 6  | -6 | -1 | 1  | -1 | 1  | -1 | 2   | -3  |
| 6  | -1 | 3  | -6 | 6  | -9 | -1 | 1  | -1 | 1  | -2  | 3   |
| 7  | 3  | -9 | 8  | -8 | 7  | -7 | -3 | 3  | -3 | 6   | 1   |
| 8  | -6 | 8  | -6 | 6  | -4 | 4  | -4 | -6 | 6  | -2  | -2  |
| 9  | 6  | -8 | 6  | -6 | 4  | -4 | 4  | -4 | -6 | 2   | 2   |
| 10 | 3  | 1  | -2 | 2  | -3 | 3  | -3 | 3  | -3 | -4  | 1   |
| 11 | -2 | 6  | -2 | 2  | 2  | -2 | 2  | -2 | 2  | -4  | -4  |

**Table 2** The M-table for the anthanthrene lattice of Fig. 1d. Note that the first (row) index is non-primed and the second (column) index is primed.

The MRR is an exact formula for conductance ratios of tight-binding representations of molecules in the weak coupling limit, when the Fermi energy is located at the centre of the HOMO-LUMO (H-L) gap. It does not depend on the size of the H-L gap and is independent of asymmetries in the contacts. In what follows, we explore the real-life implications of the MRR by evaluating the conductance ratio of two molecules both experimentally and using density functional theory (DFT) combined with non-equilibrium Green's functions.

To aid the experimental investigation of the MRR, it is helpful to select two molecules exhibiting constructive QI with very different values of $M_{ii'}$ and therefore, based on the M-table of Table 2, we compared the conductance of molecule **1**, derived from an anthanthrene core as shown in Scheme 1, with an MI of $M_{15'} = -1$, with that of the corresponding molecule **2**, for which $M_{72'} = -9$. This means that the MRR prediction for the electrical conductance of the core of **2** is $(9)^2 = 81$ times higher than that of the core of **1**. Below we demonstrate that even though **1** and **2** differ from the idealisation of Fig. 1d, this ratio is reflected in employed a mechanically controllable break junction (MCBJ) measurements of their conductances, which reveal that the single-molecule conductance of short-axis contacted anthanthrene **2** is approximately 79 times higher than that of its long-axis contacted analogue **1**.

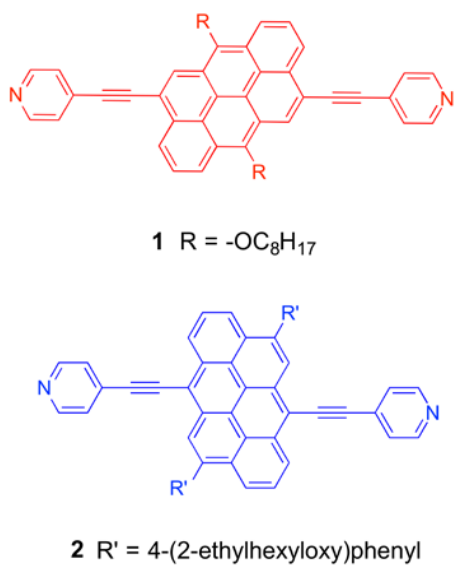

**Scheme 1** Two molecules studied experimentally, each with the anthanthrene core. Following the numbering convention in Fig. 1d, **1** is long-axis contacted with connection sites 1,5′ and **2** is short-axis contacted with connection sites 7,2′.

Anthanthrene is the compact dibenzo[*def,mno*]chrysene molecule, which together with its angular counterpart, dibenzo[*b,def*]chrysene, represents a promising building block for many applications in the field of organic electronic materials[39-41]. Advantageously, what sets these prototypical nonlinear PAHs apart from the linearly fused acenes, such as anthracene and pentacene, is the enhanced stability towards degradative chemical reactions and photooxidation[42-44]. The synthetic approach to the two novel pyridine-terminated anthanthrene derivatives (**1** and **2**, as shown in Scheme 1) is reported in the SI. To measure their single-molecule electrical conductances, we employed a MCBJ setup capable of operating in solution. In a MCBJ experiment, molecular junctions are formed by opening and closing a nanogap between two gold electrodes. For further details of conductance measurements we refer to our previous publications[22,45,46].

Fig. 2a displays typical conductance (*G*) versus distance (*Δz*) stretching traces, as plotted on a semi-logarithmic scale, and recorded for 0.1 mM molecules **1** and **2** in a solution of mesitylene and THF (4:1 v/v) using the MCBJ technique. For reference, we also plotted two traces (black curves) representing the molecule-free solution, which reveal classical tunnelling characteristics, *i.e.* an exponential decrease of the conductance upon junction elongation. After the Au-Au contacts break, the formation of molecular junctions is signalled by the presence of additional plateaus in the range $10^{-3} G_0 \geq G \geq 10^{-7.0} G_0$ ($G_0 = 2e^2/h$, quantum conductance). Typically 1000 individual conductance versus relative displacement traces (*G* vs *Δz*) were recorded for both molecules **1** and **2**, and analysed further by constructing all-data-point histograms without any data selection to extract statistically significant results from the different junction configurations (as shown in Fig. 2b). The prominent peaks between $10^{-7} G_0 < G < 10^{-4} G_0$ represent molecular junction features. The statistically-most-probable conductance of each molecular junction is obtained by fitting Gaussians to the characteristic maxima in the one-dimensional (1D) conductance histograms. As shown in Fig. 2b, the most probable conductance for the anthanthrene molecules is $10^{-4.6} G_0$ for **2** and $10^{-6.1} G_0$ for **1**, indicating that the conductance of molecule **2** is a factor of 32 higher than the conductance of **1**. However, it should be noted that the most probable conductance results from the molecular conductances associated with different contact configurations and a variety of electrode separations. To facilitate comparison with theory, it is of interest to explore the molecular conductance through fully stretched junctions, for which contact occurs via the pyridyl groups. Quantitative analyses of 2D histograms (Fig. 2c,d) reveals the evolution of molecular orientations and junction configurations during the stretching process.

The statistically averaged conductance−distance traces[44] (Fig. 2c,d) exhibit "through-space" tunnelling at the beginning of the stretching process (< 0.3 nm) and then a clear molecular plateau with slightly different conductance decays for both molecules. The analysis of stability and junction formation probability was performed by constructing the stretching distance distribution[44] shown in the inset of Fig. 2c,d. The single peak distribution suggests the junction formation probability of the anthanthrene-based molecules could reach up to ~100%. The single maximum in the plateau-length histogram represents the most probable relative characteristic stretching distance $Δz^* = 1.7$ nm for **1**, and 1.5 nm for **2**.

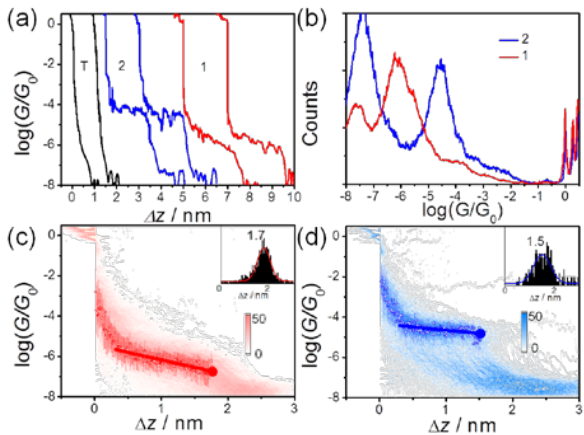

**Figure 2** (a) Individual conductance-distance traces of **1** (red) and **2** (blue) using THF/mesitylene; (b) conductance histograms of **1** (red) and **2** (blue), the sharp peak around $10^{-7.5}$ $G_0$ is attributed to the noise limit of our MCBJ setup under the current experimental condition; (c,d) 2D conductance histograms of **1** (c) and **2** (d) with statistically averaged conductance−distance traces (circles) with variation indicated by the standard deviation (bar) and linear fitting (line).[44] The solid circles represent the last data point in the linear fitting before junction rupture, and the solid error bar was determined from the Gaussian fitting of the log G peak of last data point.[44] Insets: Stretching distance distributions determined from 0.1 $G_0$ to $10^{-7}$ $G_0$ (c) and from 0.1 $G_0$ to $10^{-5.9}$ $G_0$ (d).

The most probable absolute displacement $z^*$ in an experimental molecular junction formed between two gold tips is obtained by adding the snap-back distance $\Delta z_{corr}$ to the relative displacement, namely $z^* = \Delta z^* + \Delta z_{corr}$. Taking into account $\Delta z_{corr} = (0.5 \pm 0.1)$ nm, the $z^*$ values are estimated to be 2.2 nm for **1** and 2.0 nm for **2**, which is quite close to the corresponding molecular length, and suggests that both molecules can be fully stretched during the break junction measurement. Thus the conductance of the fully-stretched molecular junction for molecules **1** and **2** are determined to be $10^{-6.7\pm0.7}$ $G_0$ (solid red circle in Fig. 2c) and $10^{-4.8\pm0.6}$ $G_0$ (solid blue circle in Fig. 2d), with the conductance ratio of ~79, which is in good agreement with the MRR. To further investigate the accuracy of the MRR and to elucidate the origins of deviations from the rule, we performed DFT-based calculations of the transmission coefficients $T(E)$ of electrons of energy $E$ passing from one electrode to the other, from which the zero temperature electrical conductance is given by Landauer formula: $G = G_0 T(E_F)$ and the room-temperature conductance obtained by integrating $T(E)$ over $E$, weighted by the derivative of the Fermi function (see methods).

Clearly the anthanthrene cores of molecules **1** and **2** do not directly contact the electrodes, but instead make indirect contact via the pyridyl rings and acetylene linkers. Therefore as an initial step, we computed the electrical conductance of the anthanthrene cores of Fig. 3a, when they are in direct contact with the gold electrodes. When the left and right electrodes are connected to atoms $i,i' = 1,5'$ respectively, this resembles the core of molecule **1**. Similarly the $i,i' = 7,2'$ connected structure resembles the core of molecule **2**. Fig. 3b shows the conductance of the anthanthrenes with *1,5'* (red curve) and *7,2'* (blue curve) connectivities obtained from a DFT-NEGF calculation, obtained in the weak coupling limit (when the gold-carbon distance is 2.4 Å). It is well known that the value of the Fermi energy predicted by DFT (*i.e.* $E^0_F = 0$ in Fig. 3b) is not necessarily reliable and therefore it is of interest to evaluate the conductance ratio for various values of $E_F$. From Fig. 3b, we find that in the range 0.2 < $E_F$ < 0.4 eV the conductance ratio varies between 69 and 88 and for a Fermi energy of $E_F$ = 0.331 eV a conductance ratio of 81 is obtained.

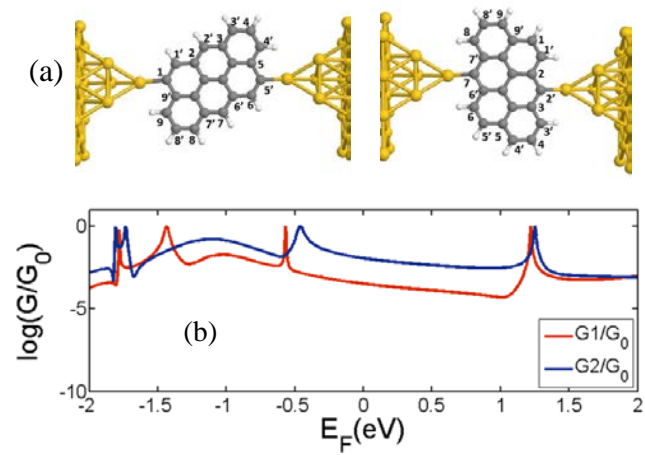

**Figure 3** (a) The anthanthrene cores connected to gold electrodes. (b) Conductance of the anthanthrene with 1,5′ (red curve) and 7,2′ (blue curve) cores obtained from DFT-NEGF.

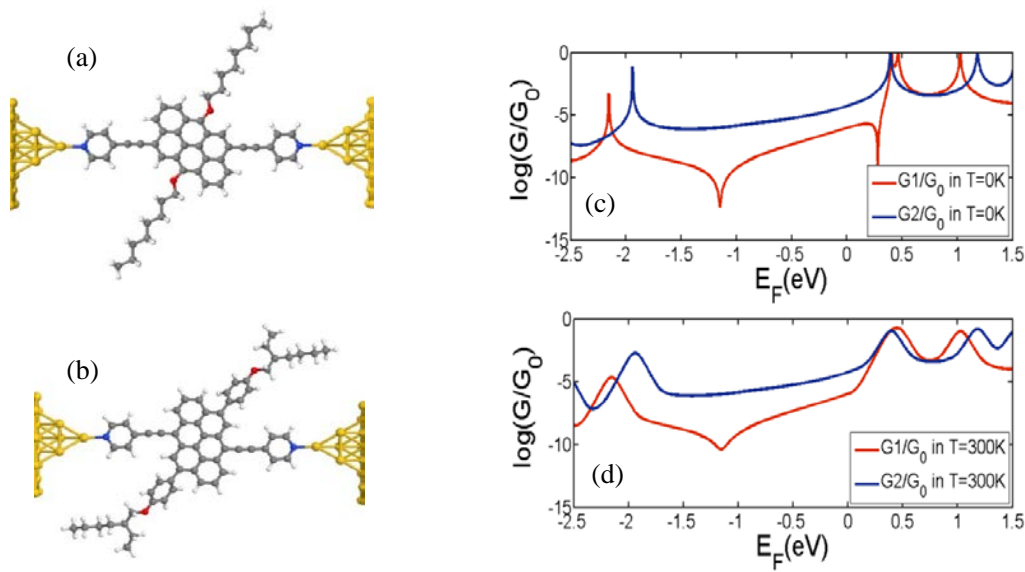

**Figure 4** (a) and (b) show the structures of **1** and **2** when the electrodes are connected to nitrogen atoms of the pyridyl anchor groups. The conductance of molecules **1** and **2** at (c) zero temperature and (d) room temperature with predicted DFT-gap from Kohn-Sham mean field Hamiltonian and with spectral adjustment based on the experimental values, respectively.

For the complete molecules measured experimentally, Fig. 4c and 4d show the logarithm of the $G/G_0$ at zero and room temperatures, respectively for molecule **1** (red solid line) and **2** (blue solid line) as a function of the Fermi energy $E_F$. Since DFT does not yield the correct H-L gap, spectral adjustment has been employed based on the experimental values of the H-L gaps[47]. As expected, Fig. 4 shows that the value of the conductance ratio depends on the location of the Fermi energy, but whatever value is chosen within the H-L gap, the conductance of **2** is much greater than that of **1**, in agreement with the MRR trend. Indeed for a value of $E_F = -0.33$ eV, the conductance of molecule **2** ($10^{-4.98}G_0$) is 81 times higher than that of molecule **1** ($10^{-6.9}\ G_0$).

Beyond the molecules investigated above, we have also examined conductances ratios of naphthalene and athracene cores obtained from the experiments reported in ref[26]. For naphthalene (molecules **4** and **6** in ref[26]) with connectivities 5,1' and 3,5' conductances of 20.8 nS and 4.1 nS were reported, which yields a measured conductance ratio of 5.1. From table 1, the MIs of these molecules are 2 and -1 respectively, yielding a MRR of 4, which is in good agreement with the experimental ratio. For anthracene (molecules **5** and **7** in ref[26]) with connectivities 6,2' and 4,7' conductances of 36.8 nS and 3.6 nS were reported, which yields a measured conductance ratio of 10.2. From the anthracene M-table presented in the SI, the MIs of these molecules are 4 and 1 respectively, yielding a MRR of 16, which also captures the trend of the experimental ratio. In this case slight disagreements may arise, because the conductance values in ref[26] include configurations in which contact is made directly with the core, rather than only through the terminal anchor groups.

## CONCLUSION

In conclusion, we have identified a new magic ratio rule (MRR), which captures the contribution of connectivity to the conductance ratios of graphene-like cores, when the coupling to the electrodes is weak and the Fermi energy coincides with the centre of the HOMO-LUMO gap. The MRR is simple to implement and exact for a tight-binding, bipartite lattice of identical sites with identical couplings, when the Fermi energy is located at the gap centre and the number of primed sites is equal to the number of unprimed sites. It states that connectivity-driven conductance ratios are simply the squares of two magic integers, whose values depend only on the connectivity to the electrodes. Based on their 'magic integers' alone, the MRR predicts that the conductance of **2** is a factor of 81 higher than that of **1**, which is in good agreement with trends obtained from both experiment and DFT calculations. Literature values of conductances for naphthalene and anthracene[26] also reveal that the MRR predicts conductance trends for these molecules. This demonstrates that connectivity is a useful starting point for designing single-molecule junctions with desirable electrical properties. As an example of such design considerations, for the purpose of connecting molecules to source-drain electrodes, a high conductance is desirable. On the other hand for the purpose of connecting to an electrostatic gate, a low conductance is needed to avoid leakage currents. Our study suggests that both features can be obtained using the same molecule provided connectivities are selected with high and low MIs for source-drain and gate electrodes, respectively.

## COMPUTATIONAL METHODS

*DFT calculations:* The optimized geometry and ground state Hamiltonian and overlap matrix elements of each structure was self-consistently obtained using the SIESTA[48] implementation of density functional theory (DFT). SIESTA employs norm-conserving pseudo-potentials to account for the core electrons and linear combinations of atomic orbitals to construct the valence states. The generalized gradient approximation (GGA) of the exchange and correlation functional is used with the Perdew-Burke-Ernzerhof parameterization (PBE)[49] a double-ζ polarized (DZP) basis set, a real-space grid defined with an equivalent energy cut-off of 250 Ry. The geometry optimization for each structure is performed to the forces smaller than 10 meV/Ang.

*Transport calculations:* The mean-field Hamiltonian obtained from the converged DFT calculation or a tight-binding Hamiltonian (using single orbital energy site per atom with Hückel parameterisation) was combined with our home-made implementation of the non-equilibrium Green's function method, GOLLUM[50], to calculate the phase-coherent, elastic scattering properties of the each system consisting of left gold (source) and right gold (drain) leads and the scattering region (molecule **1** or **2**). The transmission coefficient $T(E)$ for electrons of energy $E$ (passing from the source to the drain) is calculated via the relation: $T(E) = Trace(\Gamma_R(E)G^R(E)\Gamma_L(E)G^{R\dagger}(E))$. In this expression, $\Gamma_{L,R}(E) = i\left(\Sigma_{L,R}(E) - \Sigma_{L,R}^{\dagger}(E)\right)$ describe the level broadening due to the coupling between left (L) and right (R) electrodes and the central scattering region, $\Sigma_{L,R}(E)$ are the retarded self-energies associated with this coupling and $G^R = (ES - H - \Sigma_L - \Sigma_R)^{-1}$ is the retarded Green's function, where $H$ is the Hamiltonian and $S$ is overlap matrix. Using obtained transmission coefficient $T(E)$, the conductance could be calculated by Landauer formula $(G = G_0 \int dE\, T(E)(-\partial f/\partial E))$ where $G_0 = 2e^2/h$ is conductance quantum.

**MATHEMATICAL METHODS**

The following derivation of the MRR involves proving the three 'ratio rules' of equs. 1,2 and 3 stated below. Fig. 5a shows an example of a structure of interest, comprising a central region 2, connected by single atoms i and j to moieties on the left and right. As noted in ref [31], the Greens function $\hat{G}_{ij}(E)$ connecting sites i and j of the structure of Fig, 5a is proportional to the de Broglie wave amplitude at j, created by an incoming electron at i and the transmission coefficient $T_{ij}(E)$ is proportional to $|\hat{G}_{ij}(E)|^2$. Consequently the ratio of two transmission coefficients corresponding to connectivities i, j and l, m is given by the following Generalised Ratio Rule (GRR):

$$T_{ij}(E)/T_{lm}(E) = |\hat{G}_{ij}(E)|^2/|\hat{G}_{lm}(E)|^2 \quad (1)$$

This ratio does not depend on details of the electrodes or anchor groups, provided these are identical for both connectivities.

Furthermore, if the coupling to moieties on the left and right are sufficiently weak, and $E$ does not coincide with an eigenvalue of the isolated central region 2, $\hat{G}_{ij}(E) \approx \hat{g}_{ij}(E)$, where $\hat{g}_{ij}(E)$ is the Greens function of the isolated central region. In this case, the ratio of two transmission coefficients is given by the following Weakly-coupled Ratio Rule (WRR):

$$T_{ij}(E)/T_{lm}(E) = |\hat{g}_{ij}(E)|^2/|\hat{g}_{lm}(E)|^2 \quad (2)$$

Finally if E is located at the centre of the H-L gap (ie $E = E_F = 0$), then for a bi-partite lattice of identical sites, with equal numbers of primed and un-primed sites, described by a tight-binding model, $\hat{g}_{ij}(0) \approx (\frac{-1}{d}) M_{ij}$.

Hence the ratio of two transmission coefficients corresponding to connectivities i, j and l, m is given by the following Magic Ratio Rule (MRR):

$$T_{ij}(0)/T_{lm}(0) = (M_{ij}/M_{lm})^2 \quad (3)$$

The derivation of these ratio rules starts by noting that, the structure of Fig 5a is mathematically equivalent to the three-component system of Fig. 5b, in which the central region 2 is connected to components 1 and 3, which at large distances from 2 take the form of crystalline, periodic leads, which extend to – infinity and + infinity respectively. Conceptually, when the coupling matrices $h_{12}$ and $h_{23}$ between these regions are set to zero, such a structure consists of a 'closed inner world' (ie an inner vector space) 2, whose Greens function $g_{22}$ (for real $E$) is Hermitian, connected to an open 'outer world' composed of 1 and 3, whose Greens function is non-Hermitian[51].

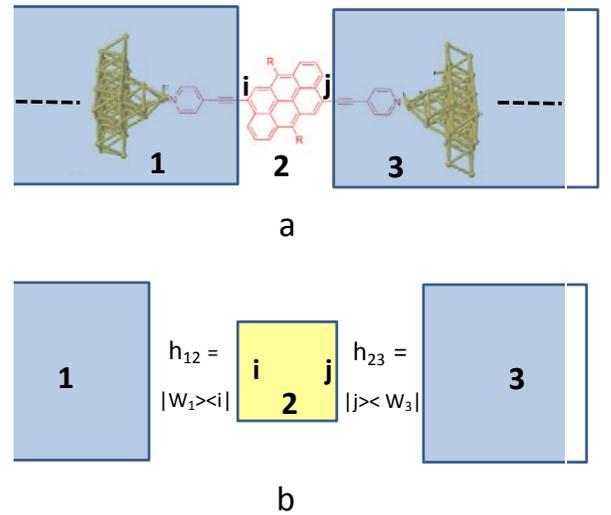

**Figure 5.** 5a shows a physical realisation of a central moiety with sites i and j connected to current-carrying bonds, which in turn are connected to anchor groups and external electrodes. 5b shows a mathematical abstraction of such a system, in which an 'inner world' 2 is connected to an 'outer world' 1 and 3 by coupling matrices $h_{12}$ and $h_{23}$.

When the coupling matrices are non-zero, the transmission coefficient $T_{ij}(E)$ from 1 to 3 is obtained from the Greens function $G_{31}$ connecting orbitals on electrode atoms of 1 to orbitals on electrode atoms of 3. In fact at large distances from 2, where $G_{31}$ can be projected onto scattering channels $|n_3>$ and $|n_1>$ of the crystalline leads of 3 and 1, the transmission coefficient can be written[51]

$$T_{ij}(E) = \sum_{n_1 n_3} T_{n_1 n_3}(E) \quad (4)$$

where $T_{n_1 n_3}(E) = V_{n_1} V_{n_3} |<n_3|G_{31}|n_1>|^2$. In this equation $V_{n_1}$ and $V_{n_3}$ are group velocities of electrons in channels $|n_1>$ and $|n_3>$. (As noted in ref [51], this expression is mathematically equivalent to the formula $T_{ij}(E) = 4 Tr\{\Gamma_1 G_{22} \Gamma_3 G_{22}^\dagger\}$, where $G_{22}$ is the Greens function of region 2, in the presence of couplings to regions 1 and 3.)

When $h_{12} = 0$ and $h_{23} = 0$, we denote the Greens functions of components 1, 2 and 3 by $g_{11}$ and $g_{22}$ and $g_{33}$ respectively. Then Dyson's equation yields

$$G_{31} = g_{33} h_{32} G_{22} h_{21} g_{11} \quad (5)$$

where

$$G_{22} = (g_{22}^{-1} - \Sigma)^{-1} \quad (6)$$

or equivalently

$$G_{22} = g_{22} + g_{22} \Sigma G_{22} \quad (7)$$

In this expression, $\Sigma = \Sigma_1 + \Sigma_3$, where $\Sigma_1 = h_{21} g_{11} h_{12}$ and $\Sigma_3 = h_{23} g_{33} h_{32}$.

So far the analysis has been rather general. We now consider the case where 1 is only coupled to a single orbital $|i>$ in 2 and 3 is coupled to only a single orbital $|j>$ in 2. (More generally, $|i>$ and $|j>$ could be arbitrary vectors in the inner vector space.) This situation is described by coupling matrices of the form $h_{21} = |W_1><i|$ and $h_{32} = |W_3><j|$, where $|W_1>$ ($|W_3>$) is a vector of matrix elements, in the space of 1 (2), describing coupling of $|i>$ ($|j>$) to orbitals in 1 (2). In this case,

$$\Sigma = \sigma_1 |i><i| + \sigma_3 |j><j|, \quad \text{where} \quad \sigma_l = <W_l|g_{ll}|W_l>, \quad (l = 1 \text{ or } 3).$$

Writing $\hat{G}_{ij} = <i|G_{22}|j>$, $\hat{g}_{ij} = <i|g_{22}|j>$ and

$$\hat{G} = \begin{pmatrix} \hat{G}_{ii} & \hat{G}_{ij} \\ \hat{G}_{ji} & \hat{G}_{jj} \end{pmatrix} \quad \text{and} \quad \hat{g} = \begin{pmatrix} \hat{g}_{ii} & \hat{g}_{ij} \\ \hat{g}_{ji} & \hat{g}_{jj} \end{pmatrix}, \text{ yields from equ. (7),}$$

$$\hat{G} = \hat{g} + \hat{g} \sigma \hat{G} \quad (8),$$

where the self-energy matrix $\sigma$ is given by $\sigma = \begin{pmatrix} \sigma_1 & 0 \\ 0 & \sigma_3 \end{pmatrix}$

Hence

$$\hat{G} = \hat{g}(1 - \sigma \hat{g})^{-1} \quad (9).$$

Similarly equ. (5) yields

$$G_{31} = g_{33} |W_3> \hat{G}_{ij} <W_1| g_{11} \quad (10).$$

This expression shows that all elements of the matrix $G_{31}$ are proportional to the single number $\hat{G}_{ij}$. Hence from equ. (4),

$$T_{ij}(E) = L(E) |\hat{G}_{ij}|^2 \quad (11),$$

which proves the GRR of equ. (1)

In equ. (11), the constant of proportionality $L(E) = \sum_{n_1 n_3} V_{n_1} V_{n_3} |<n_3|g_{33}|W_3> \hat{G}_{ij} <W_1|g_{11}|n_1>|^2$ is independent of the choice of i, j. Furthermore in equ. (9), $\hat{g}$ is independent of the couplings $|W_1>$ and $|W_3>$. On the other hand, the self-energies $\sigma_1$ and $\sigma_3$ do depend on the couplings and on i, j. However in the weak coupling limit, these vanish and therefore in equ. (9), for sufficiently-weak couplings, it is safe to neglect the product $\sigma \hat{g}$, provided $\hat{g}$ is finite. Since $\hat{g}$ is the Greens function of the isolated region 2, which diverges when E coincides with an eigenvalue of 2, this condition requires that E should lie in an energy gap of 2. (It is interesting to note that this is the opposite of the condition for applicability of the Breit-Wigner formula for resonant transmission, which requires that E should be close to an energy level of 2.) When these conditions are satisfied, $\hat{G}_{ij}(E) \approx \hat{g}_{ij}(E)$, and the WRR of equ. (2) is obtained. The WRR can be utilised by noting that $g_{22}(E) = (E - H)^{-1}$, where H is the Hamiltonian for the isolated region 2. The WRR is a generally valid whenever $\sigma \hat{g}$ can be neglected compared with unity. Physically this means that if δ is the smaller of $|E_F - E_{HOMO}|$ and $|E_F - E_{LUMO}|$, then the level broadening Γ should be much less than δ, so the ratio Γ/δ << 1.

The MRR of equ. (3) follows from the fact that if region 2 is a bi-partite lattice, then the Hamiltonian H for the isolated region 2 is of the form

$$H = \begin{pmatrix} 0 & C \\ C^t & 0 \end{pmatrix} \quad (12),$$

To obtain the transmission coefficient at the centre of the HOMO-LUMO gap, we evaluate the associated Green's function at $E = 0$, which yields

$$g_{22}(0) = \left(\frac{-1}{d}\right)\begin{pmatrix} 0 & M^t \\ M & 0 \end{pmatrix} \quad (13).$$

where $d$ is the determinant of $C$ and the matrix of MIs $M$ is the transpose of the cofactor matrix of $C$. Since the ratio of two matrix elements of $g_{22}(0)$ does not involve $d$, this completes the derivation of the MRR of equ. (3)

The condition that $g_{22}(0)$ is finite requires that $d$ should not vanish. Clearly $d = 0$ when the rows or columns of $C$ are linearly dependent, which occurs when $C$ is not a square matrix; ie when the number of primed sites is not equal to the number of un-primed sites. In this case, a transmission resonance occurs at $E = 0$ and the Breit-Wigner formula should be used. For this reason, the MRR is restricted to bi-partite lattices of identical atoms with equal numbers of primed and un-primed atoms. If this condition is not satisfied, then for non-zero energies, the WRR should be used.

Since the upper left (lower right) blocks of $g_{22}$ correspond to matrix elements between primed and primed (unprimed and unprimed) sites, the conductance vanishes when both electrodes connect to primed sites only (or unprimed sites only). For this reason, in addition to the non-trivial MIs shown in the M-tables tables, we assign an MI of zero to connectivities between primed and primed (or unprimed and unprimed) sites.

The above derivation also reveals that in addition to the MIs, each lattice possesses a second integer $d$. To each magic integer $M_{ii'}$, we assign a magic number (MN) defined by $m_{ii'} = M_{ii'}/d$. These allow the prediction of conductance ratios of molecules with different central cores via the following 'magic ratio rule' (MRR), which states that "the ratio of conductances of two molecules is equal to the squares of the ratios of their magic numbers."

Finally it is worth noting that knowledge of $T_{ij}(E)$ at $E = 0$ is particularly useful for bipartite lattices, because, $g_{22}(E)$ is symmetric about $E = 0$, so in the weak coupling limit $T_{ij}(E)$ will have a maximum or minimum (depending on the sign of the MI) at $E = 0$. Therefore at $E = 0$, $\frac{dT_{ij}(E)}{dE}$ is zero and $T_{ij}(E)$ varies slowly with $E$. Finally we note that magic numbers are a useful concept for non-bipartite lattices of identical atoms, provided det $H$ is non-zero. In this case, MIs are obtained by equating $H$ to a connectivity matrix, which contains unit matrix elements $H_{ij} = 1$ between connected sites $i$ and $j$ only and defining $M = (\det H) H^{-1}$. However in this case the spectrum is not necessarily symmetric about the gap centre and $T_{ij}(E)$ will not necessarily be either a maximum or a minimum at $E = 0$.

## ASSOCIATED CONTENT

**Supporting Information**
Experimental procedures as well as the additional computational data. This material is available free of charge via the Internet at http://pubs.acs.org.

## AUTHOR INFORMATION

**Corresponding Authors**
liu@dcb.unibe.ch
c.lambert@lancaster.ac.uk
hong@dcb.unibe.ch
s.sangtarash@lancaster.ac.uk

**Author contributions**
Y.G., S.S., C.H. and H.S. contributed equally to the paper.

**Notes**
The authors declare no competing financial interests.

<>
## Acknowledgements

This work was supported by the Swiss National Science Foundation (grant no. 200021-147143), the European Commission (EC) FP7 ITN "MOLESCO" project no. 606728 and UK EPSRC, (grant nos. EP/K001507/1, EP/J014753/1, EP/H035818/1).

## References

(1)    Chen, F.; Tao, N. *Acc. Chem. Res.* **2009**, *42*, 429.
(2)    Pisula, W.; Feng, X.; Müllen, K. *Chem. Mater.* **2010**, *23*, 554.
(3)    Carroll, R. L.; Gorman, C. B. *Angew. Chem. Int. Ed.* **2002**, *41*, 4378.
(4)    Wu, J.; Pisula, W.; Müllen, K. *Chem. Rev.* **2007**, *107*, 718.
(5)    Coskun, A.; Spruell, J. M.; Barin, G.; Dichtel, W. R.; Flood, A. H.; Botros, Y. Y.; Stoddart, J. F. *Chem. Soc. Rev.* **2012**, *41*, 4827.
(6)    Zhang, L.; Fonari, A.; Liu, Y.; Hoyt, A.-L. M.; Lee, H.; Granger, D.; Parkin, S.; Russell, T. P.; Anthony, J. E.; Brédas, J.-L. Coropceanu V.; and Briseno, A.L., *J. Am. Chem. Soc.* **2014**, *136*, 9248.
(7)    Cai, J.; Ruffieux, P.; Jaafar, R.; Bieri, M.; Braun, T.; Blankenburg, S.; Muoth, M.; Seitsonen, A. P.; Saleh, M.; Feng, X. *Nature* **2010**, *466*, 470.
(8)    Ruffieux, P.; Cai, J.; Plumb, N. C.; Patthey, L.; Prezzi, D.; Ferretti, A.; Molinari, E.; Feng, X.; Müllen, K.; Pignedoli, C. A. *ACS Nano* **2012**, *6*, 6930.
(9)    Cai, J.; Pignedoli, C. A.; Talirz, L.; Ruffieux, P.; Söde, H.; Liang, L.; Meunier, V.; Berger, R.; Li, R.; Feng, X. *Nat. Nanotechnol.* **2014**, *9*, 896.
(10)   Sedghi, G.; García-Suárez, V. M.; Esdaile, L. J.; Anderson, H. L.; Lambert, C. J.; Martín, S.; Bethell, D.; Higgins, S. J.; Elliott, M.; Bennett, N. *Nat. Nanotechnol.* **2011**, *6*, 517.
(11)   Zhao, X.; Huang, C.; Gulcur, M.; Batsanov, A. S.;

Table of Contents artwork

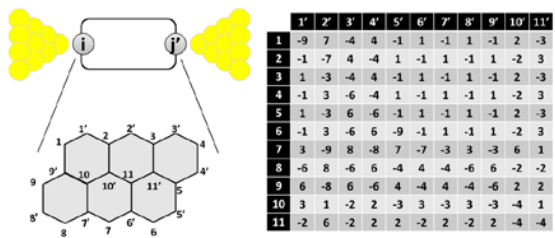

# Magic ratios for connectivity-driven electrical conductance of graphene-like molecules

Yan Geng,[1‡] Sara Sangtarash,*[,2‡] Cancan Huang,[1‡] Hatef Sadeghi,[2‡] Yongchun Fu,[1] Wenjing Hong,*[,1] Thomas Wandlowski,[1] Silvio Decurtins,[1] Colin J. Lambert,*[,2] and Shi-Xia Liu*[,1]

[1] Department of Chemistry and Biochemistry, University of Bern, Freiestrasse 3, CH-3012 Bern, Switzerland
[2] Lancaster Quantum Technology Centre, Department of Physics, Lancaster University, LA14YB Lancaster, UK
‡ Authors with equal contribution to the paper.

*Corresponding authors, e-mail: hong@dcb.unibe.ch; c.lambert@lancaster.ac.uk; liu@dcb.unibe.ch

**Synthesis and characterisation**

The synthetic routes of the anthanthrene derivatives **1** and **2** are shown in Scheme S1. Due to the poor solubility of 4,10-dibromoanthanthrone in common organic solvents, the introduction of solubilising groups at the 4- and 10-positions in the first step of the synthesis has been proven very crucial for the characterization and purification of the desired products.[R1] For instance, the key precursor **3** with solubilising groups (R = octyloxyl) was prepared via the reduction-alkylation sequence on 4,10-dibromoanthanthrone, as previously reported by Morin *et al*.[R1] It can further undergo a twofold Sonogashira coupling with 4-ethynylpyridine under standard conditions to afford compound **1** as an orange solid in 75% yield. In contrast, a twofold Suzuki coupling methodology has been applied to the preparation of the key precursor **4** functionalised with the bulky groups R' (4-(2-ethylhexyloxy)phenyl). It is then subjected to double nucleophilic attack of the lithium triisopropylsilylacetylide followed by reduction with aqueous $SnCl_2$ to produce compound **5** in 23% yield. It appears that the transformation of the quinine to the acetylene groups via the alkynylation/reduction sequence is usually accompanied with the formation of various side products leading to poor yields. The subsequent deprotection and a twofold Sonogashira coupling with 4-iodopyridine afforded compound **2** in 51% yield. All target dyes and intermediates have been fully characterized. Their NMR spectroscopic and high-resolution mass spectrometric data are consistent with their proposed structures.

**General**

Air and/or water-sensitive reactions were conducted under Ar in dry, freshly distilled solvents. Melting points were performed and not corrected. $^1$H and $^{13}$C NMR spectra were recorded on a Bruker Avance 300 spectrometer at 300 MHz and 75.5 MHz, respectively. Chemical shifts are reported in parts per million (ppm) and are referenced to the residual solvent peak (THF-$d_8$, $^1$H = 3.58 ppm, $^{13}$C = 67.57 ppm; $CD_2Cl_2$, $^1$H = 5.32 ppm, $^{13}$C = 54.00 ppm; $CDCl_3$, $^1$H = 7.26 ppm, $^{13}$C = 77.16 ppm). Coupling constants (*J*) are given in hertz (Hz) and are quoted to the nearest 0.5 Hz. Peak multiplicities are described in the following way: s, singlet; d, doublet; t, triplet; m, multiplet. FT-IR spectra were recorded on a Perkin-Elmer One FT-IR spectrometer. HRMS data was obtained with ESI (electrospray ionization) mode. Optical absorption spectra were recorded on a Perkin Elmer Lambda 900 UV/vis/NIR spectrometer. Emission spectra were measured on a Perkin Elmer LS50B luminescence spectrometer.

**Materials**

4,10-Dibromo-6,12-bis(octyloxy)anthanthrene (**3**) was prepared according to the literature procedure.[R1] 4,10-Dibromoanthanthrone was a courtesy from Heubach GmbH as Monolite Red 316801 product. Tetrahydrofuran (THF) used for anhydrous reactions was stirred over sodium/benzophenone and then distilled immediately prior to use. Unless stated otherwise, all other reagents were purchased from commercial sources and used without additional purification.

**4,10-Bis(4-pyridylethynyl)-6,12-bis(octyloxyl)anthanthrene (1).** Compound **3** (69 mg, 0.1 mmol), 4-ethynylpyridine hydrochloride (56 mg, 0.4 mmol) and $Pd(PPh_3)_4$ (12 mg, 0.01 mmol) were mixed in a solution of $Et_3N$ (6 mL) and dry THF (6 mL). After degassing for 15 minutes, the mixture was stirred at 65 °C for 24 h. After cooling to room temperature, the solvent was removed in vacuo. The crude product was purified by column chromatography on silica gel initially eluting with EtOAc and then with THF to give **1** as an orange solid. Yield: 55 mg (75%); m.p. 190-192 °C; IR (KBr): $\tilde{\nu}$ = 2210, 1593, 1536, 1469, 1342, 1285, 1186, 1094, 892, 815, 762 cm$^{-1}$; $^1$H NMR (300 MHz, THF-$d_8$): δ 8.89 (d, *J* = 7.9 Hz, 2H), 8.85-8.81 (m, 4H), 8.69 (d, *J* = 5.9 Hz, 4H), 8.30 (t, *J* = 7.4 Hz, 2H), 7.66 (d, *J* = 5.9 Hz, 4H), 4.45 (t, *J* = 6.5 Hz, 4H), 2.24-2.19 (m, 4H), 1.85-1.81 (m, 4H), 1.63-1.30 (m, 16H), 0.92 (t, *J* = 7.0 Hz, 6H); $^{13}$C NMR (75.5 MHz, THF-$d_8$): δ 151.5, 151.2, 131.8, 131.5, 129.3, 127.3, 126.8, 126.4, 125.5, 124.5, 122.7, 122.2, 120.8, 120.3, 92.8, 92.7, 77.9, 33.1, 31.9, 30.8, 30.6, 27.4, 23.8, 14.6; HRMS (ESI): *m/z* calcd for $C_{52}H_{51}N_2O_2$: 735.3945; found: 735.3947(M+H$^+$).

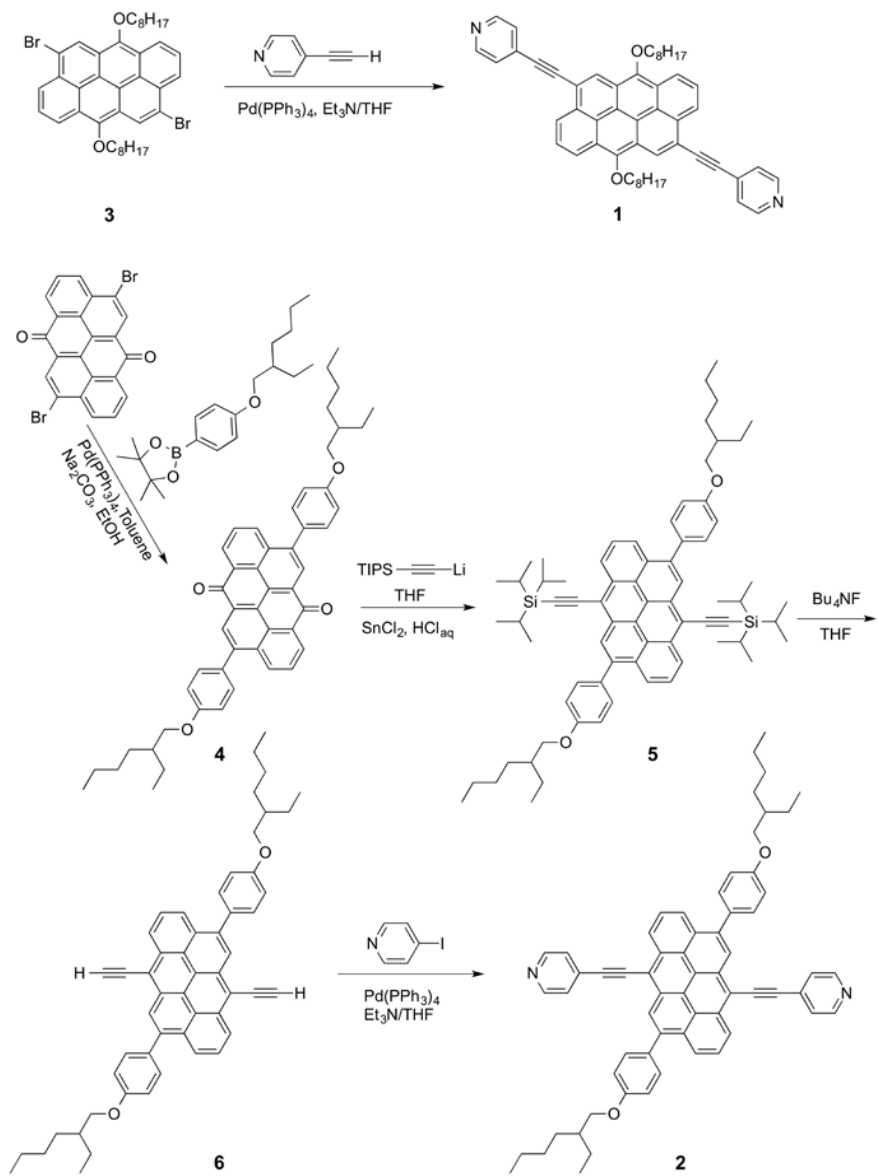

**Scheme S1** Synthetic routes to anthanthrene compounds **1** and **2**.

**4,10-Bis(4-(2-ethylhexyloxy)phenyl)anthanthrone (4).** A mixture of 4,10-dibromoanthanthrone (1.85g, 4 mmol), 4-(2-ethylhexyloxy)phenyl-4,4,5,5-tetramethyl-1,3,2-dioxaborolane (3.32 g, 10 mmol), Pd(PPh$_3$)$_4$ (0.2 g, 0.18 mmol), 18-crown-6 (0.02 g, 0.07 mmol), toluene (250 mL), EtOH (25 mL) and aqueous potassium carbonate (2 M, 40 mL) was placed in a round-bottomed flask under vacuum and then backfilled with N$_2$ three times before refluxing at 105 ºC for 20 h. After cooling to room temperature, the two layers were separated. The aqueous phase was extracted with CH$_2$Cl$_2$ (3 × 100 mL), and the combined organic phase was washed with brine (2 × 200 mL), dried over MgSO$_4$ and filtered. All solvents were removed under reduced pressure, the resultant residue was purified by column chromatography on silica gel using dichloromethane:hexane = 2:1 (v/v) as eluent to afford **4** as a red solid. Yield: 2.0 g (70%); m.p. 313-315 °C; IR (KBr): $\tilde{v}$ = 2924, 1652, 1606, 1581, 1503, 1467, 1382, 1272, 1241, 1175, 1032, 940, 838, 765, 597 cm$^{-1}$; $^1$H NMR (300 MHz, CD$_2$Cl$_2$): δ 8.53 (d, $J$ = 7.2 Hz, 2H), 8.32 (d, $J$ = 8.3 Hz, 2H), 8.24 (s, 2H), 7.72 (t, $J$ = 7.8 Hz, 2H), 7.52 (d, $J$ = 8.6 Hz, 4H), 7.12 (d, $J$ = 8.6 Hz, 4H), 4.00 (d, $J$ = 5.8 Hz, 4H), 1.85-1.79 (m, 2H), 1.63-1.26 (m, 16H), 1.03-0.93 (m, 12H); $^{13}$C NMR (300 MHz, CD$_2$Cl$_2$): δ 160.2, 143.2, 134.4, 133.4, 132.1, 131.7, 130.7, 129.6, 129.0, 127.0, 126.9, 125.0, 115.1, 71.3, 40.1, 31.1, 29.7, 24.5, 23.7, 14.5, 11.5; HRMS (ESI): $m/z$ calcd for C$_{50}$H$_{51}$O$_4$: 715.3782; found: 715.3781(M+H$^+$).

**4,10-Bis(4-(2-ethylhexyloxy)phenyl)-6,12-bis(triisopropylsilylethynyl)anthanthrene (5).** Under nitrogen atmosphere, triisopropylsilylacetylene (2.6 g, 15 mmol) was dissolved in dry THF (200 mL). At 0 °C, *n*-BuLi (2.5 M in hexanes, 5.3 mL, 13.5 mmol) was added dropwise and the resultant solution was stirred for 1.5 h at 0 °C. Compound **4** (1.06 g, 1.5 mmol) was added at 0 °C, and then the mixture was allowed to warm to room temperature and stirred overnight. SnCl$_2$·2H$_2$O (1.37g, 6 mmol) and aqueous HCl (3

M, 7 mL) were added, and the mixture was stirred for an additional 5 h. Water (50 mL) was added to quench the reaction while the organic layer was separated. The aqueous phase was extracted with $CH_2Cl_2$ (3×50 mL). The combined organic phase was dried over $MgSO_4$ and filtrated. After removing all solvents by rotavapor, the residue was purified by column chromatography on silica gel using dichloromethane:hexane = 1:4 (v/v) as eluent to give **5** as an orange solid. Yield: 0.36 g (23%); m.p. 225-227 °C; IR (KBr): $\tilde{v}$ = 2135, 1608, 1506, 1460, 1384, 1282, 1242, 1173, 994, 882, 834, 766, 738, 700, 677, 583 cm$^{-1}$; $^1$H NMR (300 MHz, CDCl$_3$): δ 9.17 (d, $J$ = 8.0 Hz, 2H), 8.83 (s, 2H), 8.46 (d, $J$ = 7.5 Hz, 2H), 8.20 (t, $J$ = 7.9 Hz, 2H), 7.74 (d, $J$ = 8.6 Hz, 4H), 7.14 (d, $J$ = 8.6 Hz, 4H), 4.01 (d, $J$ = 5.8 Hz, 4H), 1.87-1.80 (m, 2H), 1.66-1.11 (m, 58H), 1.11-0.90 (m, 12H); $^{13}$C NMR (75.5 MHz, CDCl$_3$): δ 159.4, 141.5, 132.9, 132.8, 131.5, 131.4, 131.2, 127.3, 126.7, 125.7, 124.7, 122.9, 121.5, 117.1, 114.7, 104.5, 104.4, 71.0, 39.7, 30.8, 29.4, 24.2, 23.3, 19.1, 14.3, 11.8, 11.4; HRMS (ESI): $m/z$ calcd for $C_{72}H_{92}O_2Si_2$: 1044.6630; found: 1044.6631 (M$^+$).

**4,10-Bis(4-(2-ethylhexyloxy)phenyl)-6,12-bis(ethynyl)anthanthrene (6).** Compound **5** (208 mg, 0.2 mmol) was dissolved in THF (40 mL). Bu$_4$NF·3H$_2$O (315 mg, 1 mmol) was added and the solution was stirred for 1.5 h at room temperature. Water (50 mL) was added while the organic layer was separated. The aqueous phase was extracted by $CH_2Cl_2$ (3×50 mL). The combined organic phase was dried over MgSO$_4$ and filtrated. After removing all solvents by rotavapor, the residue was purified by column chromatography on silica gel using dichloromethane:hexane = 1:4 (v/v) as eluent to give **6** as a light red solid. Yield: 125 mg (85%); > 220 °C decomposed; IR (KBr): $\tilde{v}$ = 1606, 1507, 1468, 1354, 1247, 1174, 1032, 888, 832, 762, 734, 619 cm$^{-1}$; $^1$H NMR (300 MHz, CDCl$_3$): δ 9.13 (d, $J$ = 8.0 Hz, 2H), 8.71 (s, 2H), 8.36 (d, $J$ = 7.4 Hz, 2H), 8.17 (t, $J$ = 7.9 Hz, 2H), 7.71 (d, $J$ = 8.6 Hz, 4H), 7.17 (d, $J$ = 8.6 Hz, 4H), 4.13 (s, 2H), 4.01 (d, $J$ = 5.7 Hz, 4H), 1.86-1.80 (m, 2H), 1.66-1.38 (m, 16H), 1.10-0.94 (m, 12H); $^{13}$C NMR (75.5 MHz, CDCl$_3$): δ 159.5, 141.9, 132.8, 132.8, 131.4, 131.3, 126.8, 126.7, 125.6, 125.0, 122.6, 121.6, 115.8, 114.7, 89.5, 81.3, 70.9, 39.7, 30.8, 29.3, 24.2, 23.3, 14.3, 11.4; HRMS (ESI): $m/z$ calcd for $C_{54}H_{52}O_2$: 732.3962; found: 732.3955 (M$^+$).

**4,10-Bis[4-(2-ethylhexyloxy)phenyl]-6,12-bis(4-pyridylethynyl)anthanthrene (2).** Compound **6** (73 mg, 0.1 mmol), 4-iodopyridine (80 mg, 0.4 mmol), and Pd(PPh$_3$)$_4$ (11 mg, 0.01 mmol), were mixed in Et$_3$N (6 mL) and dry THF (6 mL). After degassing for 15 minutes, the mixture was stirred at 65 °C for 24 h. After removing the solvent by rotavapor, the residue was purified by column chromatography on silica gel initially eluting with EtOAc and then with THF to give **2** as a dark red solid. Yield: 45 mg (51%); m.p.: 272-274 °C; IR (KBr): $\tilde{v}$ = 2189, 1606, 1588, 1507, 1466, 1281, 1244, 1174, 1031, 812, 762, 729, 592 cm$^{-1}$; $^1$H NMR (300 MHz, THF-$d_8$): δ 8.95 (d, $J$ = 7.9 Hz, 2H), 8.66 (d, $J$ = 5.8 Hz, 4H), 8.62 (s, 2H), 8.27 (d, $J$ = 7.4 Hz, 2H), 8.00 (t, $J$ = 7.9 Hz, 2H), 7.70 (d, $J$ = 8.6 Hz, 4H), 7.60 (d, $J$ = 5.8 Hz, 4H), 7.18 (d, $J$ = 8.6 Hz, 4H), 4.04 (d, $J$ = 5.5 Hz, 4H), 1.88-1.82 (m, 2H), 1.68-1.35 (m, 16H), 1.08-0.97 (m, 12H); $^{13}$C NMR (75.5 MHz, THF-$d_8$): δ 160.7, 151.3, 143.0, 133.5, 133.1, 132.2, 132.0, 131.8, 127.6, 127.1, 126.2, 126.1, 123.3, 122.3, 116.5, 115.6, 100.5, 92.0, 71.4, 40.9, 31.8, 30.3, 25.1, 24.2, 14.7, 11.8; HRMS (ESI): $m/z$ calcd for $C_{64}H_{59}N_2O_2$: 887.4571; found: 887.4579 (M+H$^+$).

**Optical and redox properties**

The UV−vis spectra of two anthanthrene compounds **1** and **2** were recorded in THF solutions as shown in Figure S1. Both of them show similar spectral features with fine vibronic bands which are characteristic of fused conjugated carbon-skeletons. They strongly absorb in the spectral region between 200 nm to 550 nm with extinction coefficients up to $10^5$ M$^{-1}$ cm$^{-1}$. Compound **1** presents the lowest energy absorption band at 470 nm whereas the same substituents at 6- and 12-positions of **2** lead to a red shift of 47 nm to a maximum at 517 nm. This observation is accounted for more effective π-conjugation through the aromatic core along the 6,12-axis than along the 4,10-axis, as reported in the analogous anthanthrene systems.[R2-R4] Moreover, both **1** and **2** exhibit strong fluorescence emission in THF solution with small Stokes shifts of 1080 cm$^{-1}$ and 550 cm$^{-1}$ respectively. The fluorescence excitation spectra compare well with the corresponding absorption profiles. According to the intersection of the absorption and emission spectra in solution, optical bandgaps of 2.59 and 2.36 eV for **1** and **2**, respectively, are estimated. This significant decrease in the bandgap on going from **1** to **2** is rationalized by a stronger intramolecular donor–acceptor interaction between the anthanthrene core and 4-pyridylethynyl groups at the 6,12 than 4,10-positions.

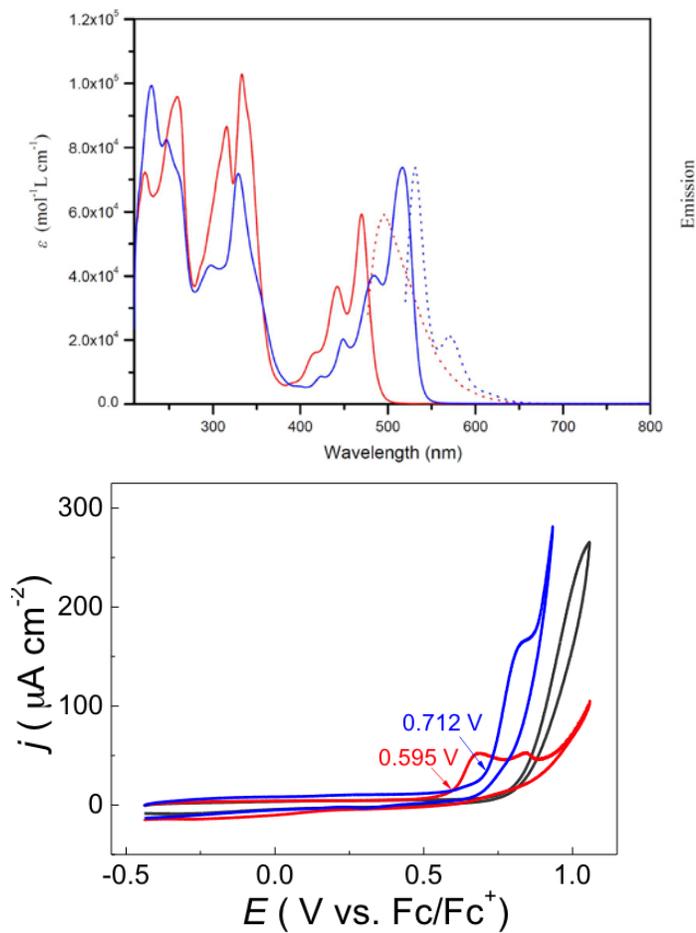

**Figure S1** (Top) UV-vis absorption (solid line) and emission (dotted line) spectra of **1** (red line) and **2** (blue line) in THF; (Bottom) Cyclic voltammograms of **1** (red line), **2** (blue line) and blank (dark-gray line) in 0.1 M Bu$_4$NPF$_6$/THF with a scan rate of 50 mV/s.

Electrochemical properties of **1** and **2** in THF were investigated by cyclic voltammetry (CV). As illustrated in Figure S1, **1** undergoes two well-separated and irreversible oxidation processes to the anthanthrene centered radical cation and dication states, In contrast, **2** only shows the first irreversible oxidation at a much higher oxidation potential than **1**. On account of more effective π-conjugation through the aromatic core along the 6,12-axis than along the 4,10-axis, the stronger electron-withdrawing effect of the 4-pyridylethynyl groups in **2** than in **1** is expected to lower the HOMO energy level and destabilize the anthanthrene radical, thus leading to a substantial positive-shift of the oxidation potential. Furthermore, the electron-donating effect of the octyloxy groups in **1** causes the anthanthrene core to be more easily oxidized compared to **2**. Based on the onset of the first oxidation potential of 0.60 V and 0.71 V for **1** and **2**, respectively, the corresponding HOMO energy levels of -5.40 eV and -5.51 eV are estimated according to the following equation: $E_{HOMO} = [-(E\text{onset} (vs \text{ ferrocene}) + 4.8)]$ eV, where 4.8 eV is the energy level of ferrocene below the vacuum level.[R5] From the optical bandgaps, the LUMO energy levels are expected to be around -2.81 eV and -2.59 eV for **1** and **2**, respectively, which are beyond the negative potential window available in THF solvent.

**Iso-surfaces obtained from DFT**

Iso-surfaces of the HOMO and LUMO for molecules **1** and **2** are shown in Fig. S2a and S2b. As indicated by the transport resonances in Fig. S3, the DFT calculation (see methods) shows that the HOMO-LUMO gap of the molecule **1** is bigger than molecule **2**, in agreement with experiment.

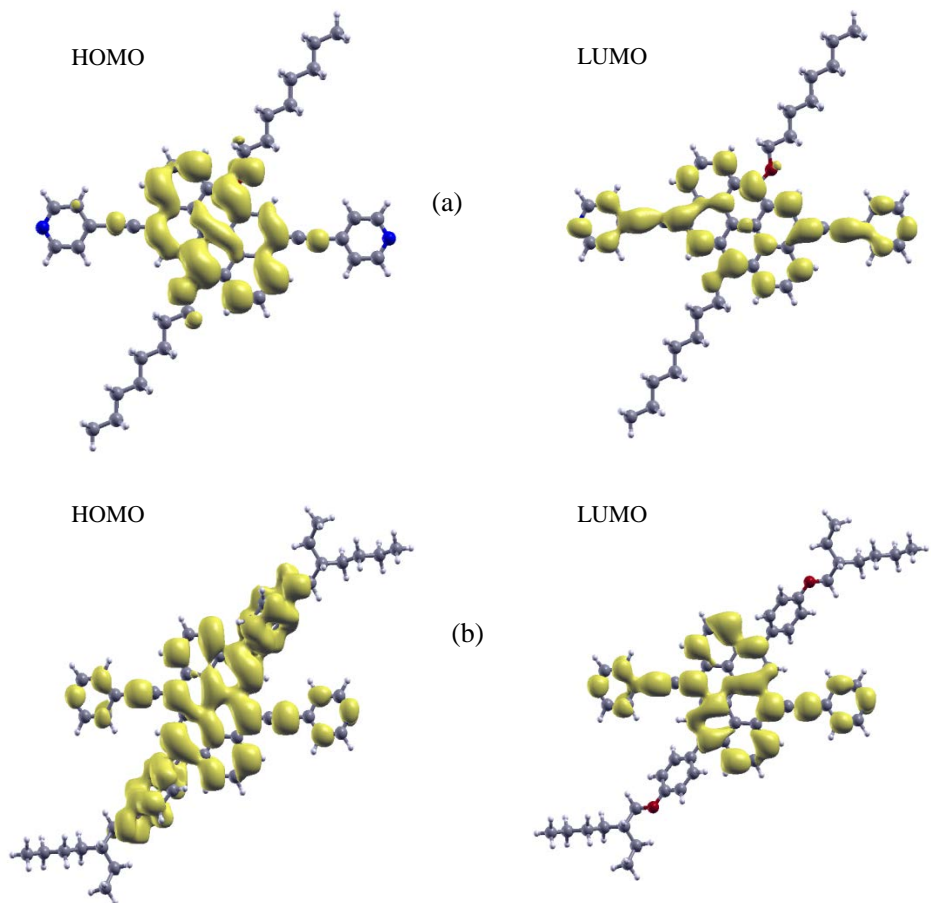

**Figure S2** Local density of state in HOMO and LUMO energy levels for molecule (a) **1** and (b) **2**.

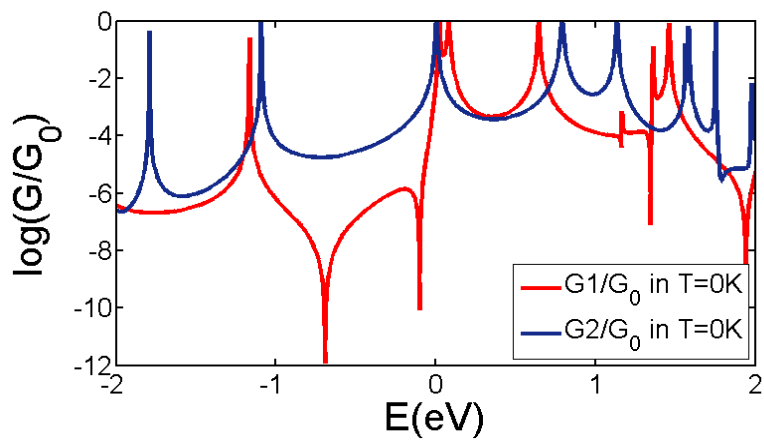

**Figure S3** The transmission coefficient for molecule **1** (red line) and **2** (blue line), before the spectral adjustment applies.

**Examples of M-tables of magic integers**

We first restrict the discussion to bipartite lattices with equal numbers of primed and non-primed sites. To obtain the M-table for a given lattice, first construct a connectivity table C, with rows labelled by primed integers and columns by unprimed integers, such that the entry $C_{i'i}$ contains a '1' if sites i' and i are connected and zero otherwise. The corresponding M-table M is then defined to be the transpose of the cofactor matrix of C. This means that if the determinant of the matrix obtained by removing the ith column and i'th row of C is denoted $d_{ii'}$, then $M_{ii'} = (-1)^{(i+i')} d_{ii'}$.

**Example 1. A benzene ring**

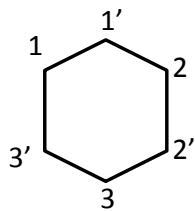

C=

| | 1 | 2 | 3 |
|---|---|---|---|
| 1' | 1 | 1 | 0 |
| 2' | 0 | 1 | 1 |
| 3' | 1 | 0 | 1 |

M=

| | 1' | 2' | 3' |
|---|---|---|---|
| 1 | 1 | -1 | 1 |
| 2 | 1 | 1 | -1 |
| 3 | -1 | 1 | 1 |

Clearly M is related to the inverse of C by M = d × $C^{-1}$, where d is the determinant of C. For this lattice $d$ = det C = 2.

**Example 2. Naphthalene**

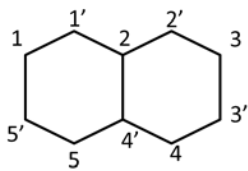

C=

| | 1 | 2 | 3 | 4 | 5 |
|---|---|---|---|---|---|
| 1' | 1 | 1 | 0 | 0 | 0 |
| 2' | 0 | 1 | 1 | 0 | 0 |
| 3' | 0 | 0 | 1 | 1 | 0 |
| 4' | 0 | 1 | 0 | 1 | 1 |
| 5' | 1 | 0 | 0 | 0 | 1 |

M=

| | 1' | 2' | 3' | 4' | 5' |
|---|---|---|---|---|---|
| 1 | -2 | 1 | -1 | 1 | -1 |
| 2 | -1 | -1 | 1 | -1 | 1 |
| 3 | 1 | -2 | -1 | 1 | -1 |
| 4 | -1 | 2 | -2 | -1 | 1 |
| 5 | 2 | -1 | 1 | -1 | -2 |

For this lattice, $d$ = det C = 3, so $C^{-1}$ = (1/3) M, as is easily verified by evaluating the product C × M.

**Example 3. Anthracene**

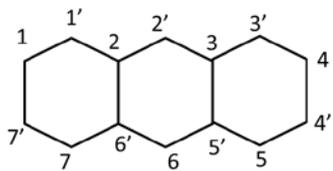

For this lattice, *d* = 4

C=

|    | 1 | 2 | 3 | 4 | 5 | 6 | 7 |
|----|---|---|---|---|---|---|---|
| 1' | 1 | 1 | 0 | 0 | 0 | 0 | 0 |
| 2' | 0 | 1 | 1 | 0 | 0 | 0 | 0 |
| 3' | 0 | 0 | 1 | 1 | 0 | 0 | 0 |
| 4' | 0 | 0 | 0 | 1 | 1 | 0 | 0 |
| 5' | 0 | 0 | 1 | 0 | 1 | 1 | 0 |
| 6' | 0 | 1 | 0 | 0 | 0 | 1 | 1 |
| 7' | 1 | 0 | 0 | 0 | 0 | 0 | 1 |

M=

|   | 1' | 2' | 3' | 4' | 5' | 6' | 7' |
|---|----|----|----|----|----|----|----|
| 1 | -3 | 2  | -1 | 1  | -1 | 1  | -1 |
| 2 | -1 | -2 | 1  | -1 | 1  | -1 | 1  |
| 3 | 1  | -2 | -1 | 1  | -1 | 1  | -1 |
| 4 | -1 | 2  | -3 | -1 | 1  | -1 | 1  |
| 5 | 1  | -2 | 3  | -3 | -1 | 1  | -1 |
| 6 | -2 | 4  | -2 | 2  | -2 | -2 | 2  |
| 7 | 3  | -2 | 1  | -1 | 1  | -1 | -3 |

**Example 4. Anthanthrene**

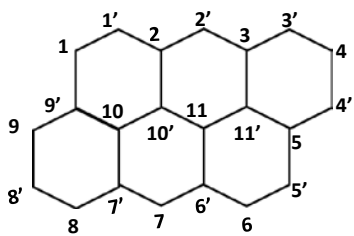

C=

|     | 1 | 2 | 3 | 4 | 5 | 6 | 7 | 8 | 9 | 10 | 11 |
|-----|---|---|---|---|---|---|---|---|---|----|----|
| 1'  | 1 | 1 | 0 | 0 | 0 | 0 | 0 | 0 | 0 | 0  | 0  |
| 2'  | 0 | 1 | 1 | 0 | 0 | 0 | 0 | 0 | 0 | 0  | 0  |
| 3'  | 0 | 0 | 1 | 1 | 0 | 0 | 0 | 0 | 0 | 0  | 0  |
| 4'  | 0 | 0 | 0 | 1 | 1 | 0 | 0 | 0 | 0 | 0  | 0  |
| 5'  | 0 | 0 | 0 | 0 | 1 | 1 | 0 | 0 | 0 | 0  | 0  |
| 6'  | 0 | 0 | 0 | 0 | 0 | 1 | 1 | 0 | 0 | 0  | 1  |
| 7'  | 0 | 0 | 0 | 0 | 0 | 0 | 1 | 1 | 0 | 1  | 0  |
| 8'  | 0 | 0 | 0 | 0 | 0 | 0 | 0 | 1 | 1 | 0  | 0  |
| 9'  | 1 | 0 | 0 | 0 | 0 | 0 | 0 | 0 | 1 | 1  | 0  |
| 10' | 0 | 1 | 0 | 0 | 0 | 0 | 0 | 0 | 0 | 1  | 1  |
| 11' | 0 | 0 | 1 | 0 | 1 | 0 | 0 | 0 | 0 | 0  | 1  |

M =

|    | 1'  | 2'  | 3'  | 4'  | 5'  | 6'  | 7'  | 8'  | 9'  | 10' | 11' |
|----|-----|-----|-----|-----|-----|-----|-----|-----|-----|-----|-----|
| 1  | -9  | 7   | -4  | 4   | -1  | 1   | -1  | 1   | -1  | 2   | -3  |
| 2  | -1  | -7  | 4   | -4  | 1   | -1  | 1   | -1  | 1   | -2  | 3   |
| 3  | 1   | -3  | -4  | 4   | -1  | 1   | -1  | 1   | -1  | 2   | -3  |
| 4  | -1  | 3   | -6  | -4  | 1   | -1  | 1   | -1  | 1   | -2  | 3   |
| 5  | 1   | -3  | 6   | -6  | -1  | 1   | -1  | 1   | -1  | 2   | -3  |
| 6  | -1  | 3   | -6  | 6   | -9  | -1  | 1   | -1  | 1   | -2  | 3   |
| 7  | 3   | -9  | 8   | -8  | 7   | -7  | -3  | 3   | -3  | 6   | 1   |
| 8  | -6  | 8   | -6  | 6   | -4  | 4   | -4  | -6  | 6   | -2  | -2  |
| 9  | 6   | -8  | 6   | -6  | 4   | -4  | 4   | -4  | -6  | 2   | 2   |
| 10 | 3   | 1   | -2  | 2   | -3  | 3   | -3  | 3   | -3  | -4  | 1   |
| 11 | -2  | 6   | -2  | 2   | 2   | -2  | 2   | -2  | 2   | -4  | -4  |

For this lattice, $d = 10$